# When Literature Data Mislead Artificial Intelligence in Materials Discovery

Qian Wang[1], Ying Li[1], Ryuhei Sato[2], Hidemi Kato[3], Shin-ichi Orimo[1,3], Hao Li[1], Eric Jianfeng Cheng[1,*]

[1] Advanced Institute for Materials Research (WPI-AIMR), Tohoku University, Sendai 980–8577, Japan

[2] Department of Materials Engineering, The University of Tokyo, Tokyo 113-8656

[3] Institute for Materials Research, Tohoku University, Sendai 980–8577, Japan

* Email: ericonium@tohoku.ac.jp

**Abstract**

Artificial intelligence (AI) increasingly treats scientific literature as a data source for building databases, training predictive models, and guiding discovery. Yet literature-derived datasets often assume that reported experimental values are internally consistent and directly reusable. Here, we analyze this assumption using solid electrolyte (SE) conductivity data as a representative materials-science case. By tracing values from source articles to curated datasets, we identify recurrent text-figure mismatches, ambiguous axis annotations, unit inconsistencies, and missing measurement context. These discrepancies are often numerically plausible and therefore difficult to detect through routine preprocessing, but they can propagate as structured label noise during database construction and machine-learning reuse. A cross-database example shows how ambiguous reporting can create a 100-fold conductivity error. Our analysis reframes data accuracy as an infrastructure requirement for artificial-intelligence-driven discovery and motivates traceable reporting, curation, and validation practices for reusable scientific data.

**Keywords:** AI for science; Data reliability; Scientific databases; Structured label noise; Literature-derived data; Materials informatics; Solid electrolytes

Artificial intelligence (AI) is increasingly transforming scientific literature from a human-readable record into a machine-readable data source for discovery. Across materials science, chemistry, and energy research, published values are now extracted, standardized, and aggregated into databases that support machine learning (ML), literature mining, automated screening, and autonomous experimentation.[1–4] This shift is part of a broader movement toward data-centric AI, where model reliability depends not only on algorithms, but also on the quality, provenance, and reusability of the data used to build them.[5–7] It also creates a quiet but consequential assumption: that literature-derived data are internally consistent, physically interpretable, and directly reusable.

This assumption is fragile because experimental data are not merely numerical labels. They are measurements embedded in a scientific context. A reported value may depend on the unit convention, measurement temperature, plotted quantity, fitting range, sample-processing history, and whether the value was directly measured, interpolated, or extrapolated. Expert readers can often reconstruct this context by reading the full paper, figures, and supporting information together. Databases and machine-learning workflows, however, require each value to be converted into an explicit, standardized and comparable entry. When the surrounding context is incomplete or inconsistent, a value may remain numerically plausible while becoming semantically ambiguous. Such entries are difficult to detect by routine outlier screening, but can propagate through curation as structured label noise rather than random error. This resembles the broader problem of data cascades in AI systems, where upstream data issues can remain hidden until they affect downstream models and decisions.[8]

Here we examine this problem using solid electrolyte (SE) conductivity data as a representative case in materials discovery. Ionic conductivity is widely used for benchmarking, database construction and ML screening, but it is also highly sensitive to reporting conventions. Several informatics efforts have sought to transform literature-reported conductivity values into reusable solid-electrolyte datasets, including expert-curated lithium-ion-conductor databases, the Dynamic Database of Solid-State Electrolytes (DDSE), and text-mined ionic-conductivity datasets.[9–12] These efforts differ in scale and curation strategy, but they converge on the same requirement: published values must be reported in a form that is sufficiently clear for standardization, comparison, and downstream reuse in computational workflows.

**Figure 1** summarizes the conceptual distinction between reliable and distorted workflows linking scientific literature to AI-assisted analysis. In an ideal workflow, experimental values are reported with sufficient contextual information to allow faithful transfer from the literature to a database and then to AI-assisted analysis (**Fig. 1A**). In a distorted workflow, ambiguity introduced at the publication stage is not eliminated by curation, but is instead transformed into uncertain database labels that may be reused by downstream models (**Fig. 1B**). By tracing values from source articles to curated datasets, we show how text-figure mismatches, ambiguous annotations, unit inconsistencies and incomplete measurement context can distort data reuse. Our aim is not only to improve reporting in SE research, but also to highlight a broader data-reliability problem in AI-enabled science.

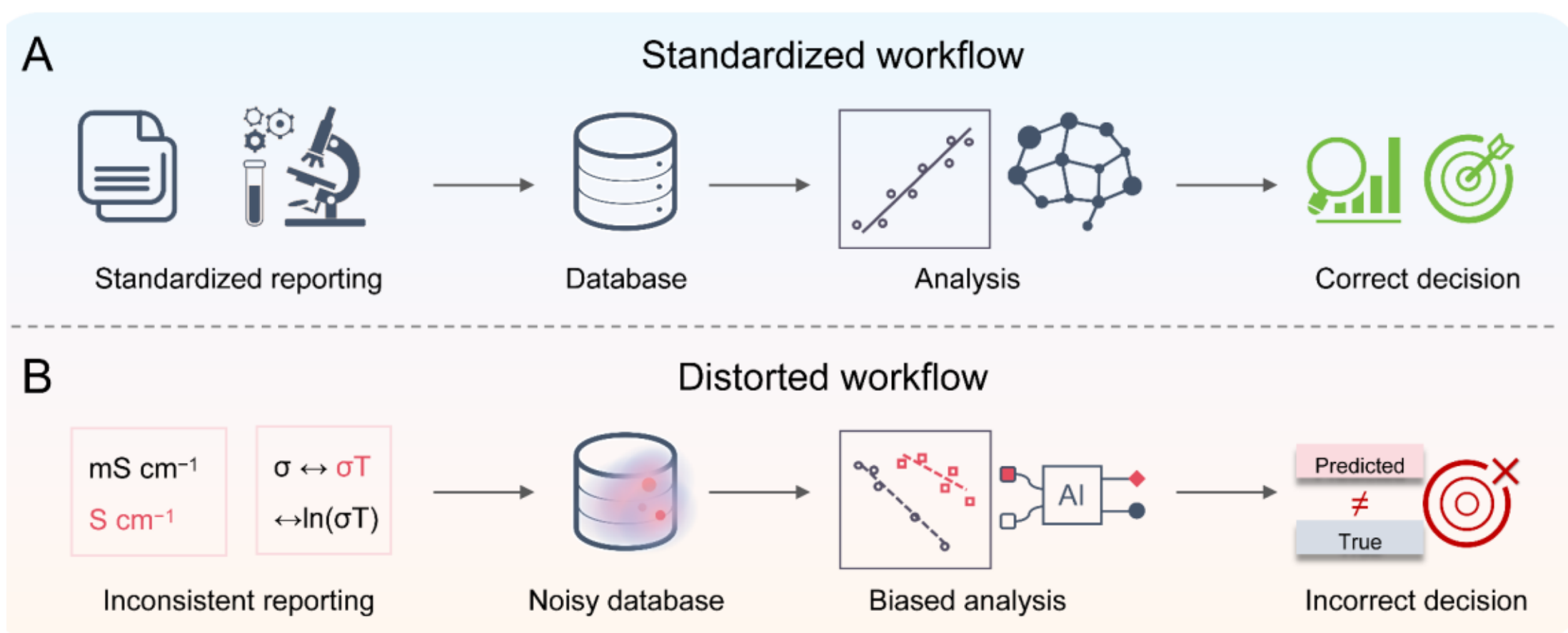

**Figure 1.** Schematic illustration of standardized (**A**) and distorted (**B**) workflows from literature to database and AI-assisted analysis.

Solid electrolyte research provides a useful test case because ionic conductivity is both widely reused and highly context-dependent. It is routinely used for benchmarking, database construction and machine-learning screening, yet its reported value can depend on temperature, unit convention, conductivity type, fitting procedure, sample processing and microstructure. This sensitivity is well documented: an interlaboratory study of thiophosphate-based SEs showed substantial variation in reported total ionic conductivity and activation energy for nominally identical samples,[13] while studies on garnet-type $Li_7La_3Zr_2O_{12}$ further illustrate that processing-dependent factors such as sintering route, densification, and microstructure can also affect ionic transport.[14,15] In the literature, conductivity-related information often requires substantial expert judgment to identify comparable experimental values and assemble a reliable machine-learning dataset.[9] During the construction of the Digital Battery Platform (*DigBat*), a large-scale database spanning both polymer and inorganic SEs, we found that the main challenge often lay not in collecting conductivity values themselves, but in determining how reported values should be interpreted and standardized. Such ambiguity can lead to repeated extraction errors even among researchers familiar with the SE field, indicating that the problem is structural rather than merely inattentive. Data quality should therefore be understood not only as a literature-level concern, but also as a database and AI issue.

## Results

### Literature ambiguity as structured label noise

Our curation experience shows that reporting problems in SE literature mainly fall into four categories: text–figure inconsistency, annotation ambiguity, unit inconsistency, and incomplete measurement context. These problems are especially important for ionic conductivity and other Arrhenius-derived quantities, including activation energy. **Figure 2** illustrates how these issues arise in practice and why they become consequential during database construction.

In representative gel polymer electrolyte cases, the conductivity value highlighted in the abstract or main text differs from the value that can be obtained by replotting the original conductivity–temperature data. As shown in **Fig. 2A**, the value highlighted in the text is $7.2\times10^{-4}$ S cm$^{-1}$ at 25 °C, whereas re-reading the original plot shows that the same value is associated with a point closer to 30 °C. At the level of an individual paper, such a discrepancy may appear minor because readers may rely primarily on the value stated in the text. During curation, however, the issue becomes immediate: should the stored room-temperature conductivity follow the quoted statement or the value-position relationship implied by the original figure? Repeated across many papers, this type of mismatch introduces systematic uncertainty into database labels rather than remaining a local presentation error.

**Fig. 2B** highlights a second type of ambiguity, in which the difficulty lies not in the plotted point itself but in the meaning of the plotted quantity. Reanalysis of the same plotted value under different plausible y-axis interpretations, such as log(σ), ln(σT), and log(σT), yields substantially different room-temperature conductivities. In the representative example, these interpretations lead to extracted values of $7.2\times10^{-4}$, $1.4\times10^{-4}$, and $2.4\times10^{-6}$ S cm$^{-1}$, respectively. A related ambiguity can also arise from the x-axis when a temperature reported as 25 °C must be converted to 298 K and then to the corresponding Arrhenius coordinate, 1000/T = 3.36 K$^{-1}$. If this conversion is not explicitly documented, room-temperature values may be read from an incorrect *x*-position or confused with extrapolated values. These ambiguities are difficult to detect during routine screening because all resulting values can remain numerically plausible within the broad conductivity range reported for solid electrolytes. This problem also extends to activation energy, because the fitted Arrhenius quantity and the selected temperature range are not always described with sufficient precision for reproducible extraction.

Unit inconsistency is especially dangerous in database construction because it can shift conductivity values by orders of magnitude without producing obvious outliers.

As shown in **Fig. 2C**, the reported room-temperature conductivity is $8.54 \times 10^{-3}$ S $cm^{-1}$ at 25 °C. If this value is interpreted directly in S $cm^{-1}$, it corresponds to ln(σT)=0.93, which does not match the plotted 25 °C position. By contrast, after conversion to 0.854 S $m^{-1}$, the corresponding value becomes ln(σT)=5.54, which agrees with the original plot. This unit-label inconsistency can lead to erroneous database standardization.

An incomplete measurement context creates a different type of ambiguity. A conductivity value may be described as "room temperature" without specifying the exact temperature, or reported without clarifying whether it was directly measured, interpolated between neighboring points, or derived from Arrhenius fitting. Although these values are often cited in the same way in the literature, they are not strictly equivalent in terms of experimental meaning and database standardization. A related issue also applies to activation energy: the temperature range and data points selected for Arrhenius fitting are not defined according to a uniform convention across studies, which limits the direct comparability of reported activation energies.

The broader manifestation of these problems during curation is summarized quantitatively in **Fig. 2D**. In the screened gel SE literature, 83 papers were examined, among which 9 cases (10.8%) were identified as text–figure inconsistency and 3 cases (3.6%) as annotation ambiguity (**Table S1**). Thus, the identifiable issues in gel SEs were mainly concentrated in discrepancies between quoted and plotted conductivity values. By contrast, the inorganic SE cases showed a more complex and heterogeneous error structure. The 16 source cases listed in **Table S2** correspond to 25 issue occurrences because several papers involved more than one type of ambiguity. Unit inconsistency was the most frequent issue, with 13 occurrences, followed by annotation ambiguity with 7 occurrences, text–figure inconsistency with 4 occurrences, and incomplete measurement context with 1 occurrence. This distribution indicates that inorganic SE curation is challenged not only by mismatches between reported and plotted values, but also by ambiguity in the plotted quantity, unit convention, and measurement context.

Case I16 further shows that such ambiguity can propagate beyond a single database. In the original source article, the plotted quantity is σT and σ is expressed in S $m^{-1}$, whereas a subsequent SE database used in a machine-learning study interpreted the value as S $cm^{-1}$, leading to a 100-fold error. This example illustrates a failure of data lineage: once an ambiguous value enters a curated dataset, downstream users may inherit the standardized number without visibility into the interpretive decision that produced it. The fact that this occurred in an SE-focused database indicates that the problem is not merely manual carelessness,[9] but a consequence of ambiguous quantity and unit reporting that remains difficult even for domain experts. This cross-database example reinforces the need for conductivity values to be reported with explicit plotted quantities, units, and conversion conventions.

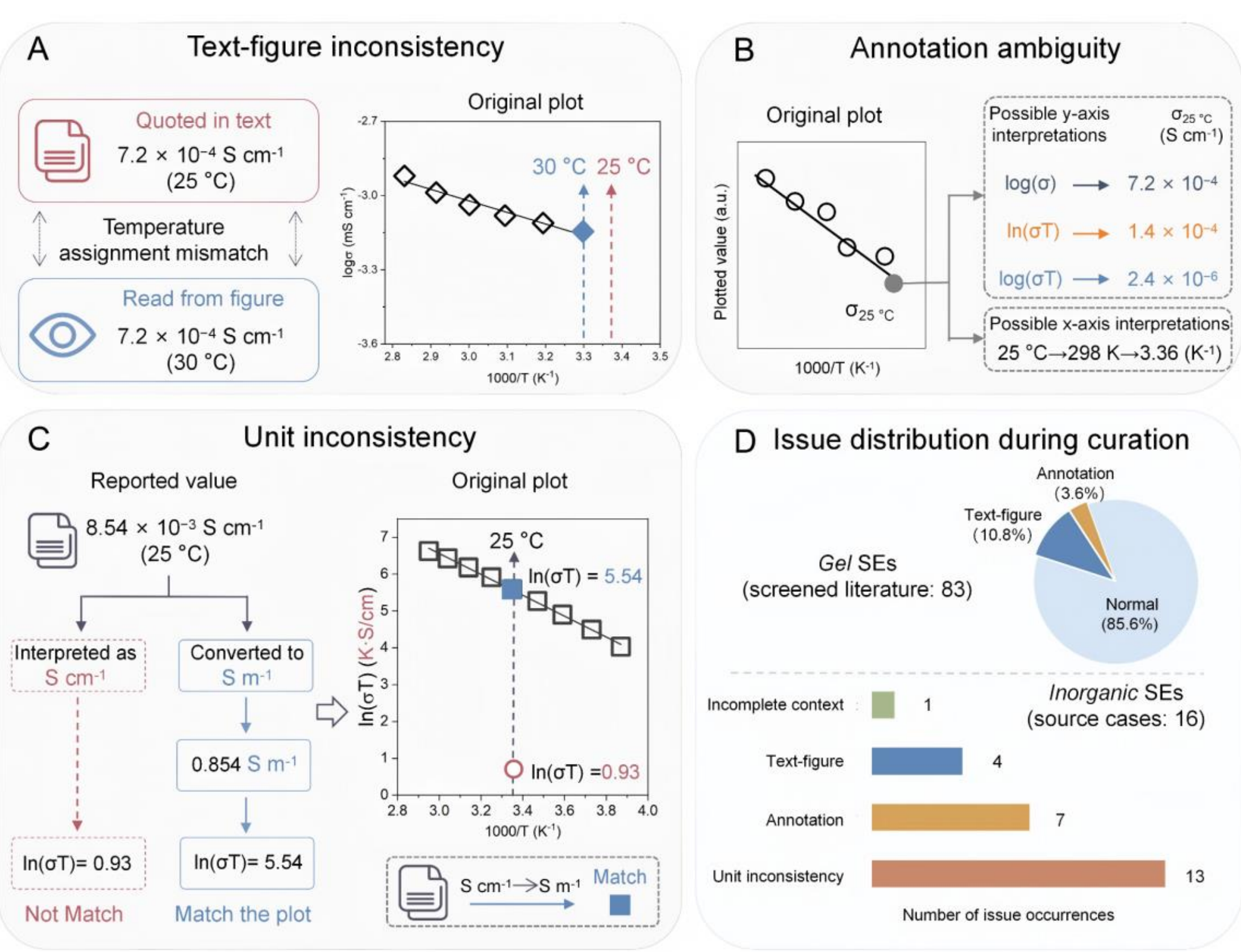

**Figure 2**. Representative case analyses of conductivity reporting ambiguity based on replotted literature data. (**A**) Comparison between quoted and replotted conductivity values (gel SE case: G5). (**B**) Conductivity interpretation under ambiguous annotation

(representative gel and inorganic SE cases). (**C**) Effect of unit inconsistency on standardized conductivity assignment (inorganic SE case: I7). (**D**) Manifestation of issue categories during curation, showing the overall distribution in screened gel SE literature and the issue-occurrence distribution among 16 inorganic SE source cases. For gel SEs, statistics are calculated at the paper level from 83 screened publications. For inorganic SEs, statistics are calculated based on issue occurrence across 16 source cases and should not be interpreted as prevalence across the full inorganic SE literature.

Taken together, the cases in **Fig. 2** show that the main difficulty in curation does not arise from missing values alone, but from the need to resolve ambiguity before a value can be stored as a standardized database entry. In this sense, curation is not a passive transcription step, but an interpretive scientific process in which the numerical meaning of reported data must first be reconstructed before the data can be standardized.

## Discussion

### Why this matters for databases and AI

Unlike random noise, these errors are structured because they arise from recurring reporting conventions, unit conversions, and axis interpretations; they may therefore bias learned correlations rather than merely increase uncertainty. For database construction, the consequence of such ambiguity is not simply inconvenience. Each ambiguous case forces a curator to decide whether a value should be included, how it should be normalized, which physical quantity it represents, and whether an uncertainty flag is needed. These decisions directly shape the comparability, traceability, and long-term reusability of the dataset.

The implications become more serious in AI workflows. Machine-learning models generally assume that training labels are internally consistent and directly comparable. In practice, however, semantically inconsistent conductivity values are often numerically plausible and therefore difficult to detect through routine preprocessing alone. Unlike obvious outliers, such entries can remain embedded in the dataset while silently introducing structured label noise. As a result, the apparent size of the dataset may increase, while its true information quality decreases. This hidden ambiguity can distort trend analysis, weaken descriptor-property relationships, and reduce the reliability of AI-based screening. More broadly, **Fig. 3A** summarizes why conductivity data accuracy matters not only for literature comparison but also for database quality and AI-assisted interpretation.

More generally, reusable experimental data require three layers of traceability: the original reported value, the physical meaning of the value, and the curation decision used to standardize it. The representative cases discussed above suggest that many curation difficulties could be reduced by a small set of clearer reporting practices. A practical way forward is to adopt a minimum reporting standard for ionic conductivity in SE research. To reduce text–figure inconsistency, key conductivity values emphasized in the text should be directly cross-checked against the plotted or tabulated data. To avoid annotation ambiguity, Arrhenius-type figures should explicitly define whether the plotted quantity is σ, log10(σ), ln(σ), σT, ln(σT), or log10(σT), and should state the conductivity unit used before any σT transformation. To avoid unit inconsistency, the same unit convention should be applied consistently across the text, figures, tables, and Supporting Information. To reduce ambiguity in the measurement context, authors should specify the exact temperature associated with each quoted value and clarify whether it is directly measured, interpolated, or fitted. For activation energy, the fitting range and the exact plotted quantity used in Arrhenius analysis should also be reported explicitly. These practical measures are summarized in **Fig. 3B** and should be regarded as minimal conditions for literature values to remain reusable on the database scale.

These measures should not be viewed merely as editorial refinements. As literature data increasingly serve as inputs to databases and AI workflows, accurate reporting becomes essential for reproducibility, database interoperability, and trustworthy data reuse.

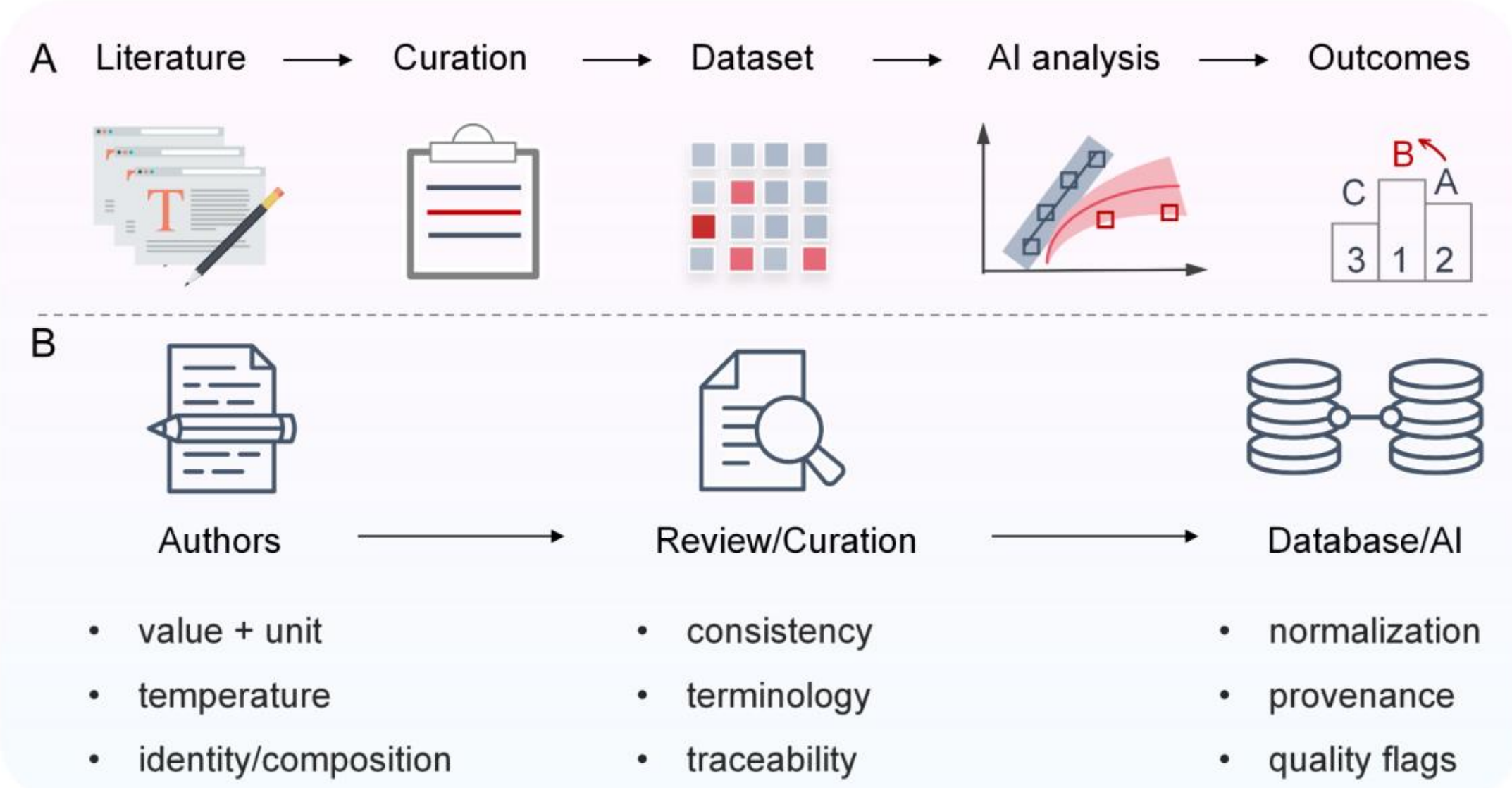

**Figure 3**. Consequences of conductivity reporting ambiguity for literature reuse, database quality, and AI-assisted analysis (**A**), and practical directions for improving conductivity reporting in SE research (**B**).

As scientific literature increasingly serves as a machine-readable substrate for databases, models, and automated discovery workflows, the reliability of reported experimental values becomes part of the infrastructure of AI-enabled materials research. Under this new role, small ambiguities in reported experimental values can be amplified during extraction, standardization, aggregation, and reuse. In SE research, ionic conductivity provides a representative example: a single numerical value may lose its physical meaning if temperature, unit, conductivity type, plotted quantity, fitting protocol, or processing context is not clearly specified.

Data accuracy should therefore be regarded not as a downstream presentation issue, but as a prerequisite for reproducible science and trustworthy AI-assisted discovery. Improving AI for materials discovery will require not only better algorithms or larger datasets, but also a more accurate, standardized, and reusable scientific record. The cross-database propagation of ambiguous values further shows that data quality cannot be guaranteed by database construction alone, even when curation is performed by domain experts. Future SE databases should therefore record not only the final standardized conductivity value, but also the original plotted quantity, unit convention, conversion procedure, measurement context, and curation decision. Such traceable records would make databases more auditable, reduce the risk of repeated errors across data infrastructures, and provide more reliable inputs for AI-driven materials discovery.

# Methods

## Data sources and selection

Literature data were collected from peer-reviewed publications on polymer and inorganic solid electrolytes and complemented by curated entries from established data resources, including the Digital Battery Platform (DigBat) and recent expert-curated lithium-ion-conductor datasets.[9] The primary dataset analyzed in this work consisted of experimentally reported ionic conductivity values together with their corresponding conductivity–temperature relationships.

To provide compositional and structural context, the collected materials were cross-referenced with widely used materials databases, including the Materials Project, AFLOW, the Open Quantum Materials Database (OQMD), and the Inorganic Crystal Structure Database (ICSD). These databases were used only as auxiliary references for checking material composition, crystal structure, phase identity, space-group information and database identifiers where available; they were not used as primary sources of experimental ionic conductivity.

Publications were selected based on the availability of conductivity–temperature relationships presented in graphical or tabulated form, enabling independent reanalysis. Only studies reporting experimentally measured ionic conductivity, primarily via impedance spectroscopy, were included in the main analysis. Computational-only data were excluded from the conductivity benchmark but were used when necessary for structural or compositional consistency checks.

In total, the curated dataset comprised 3,814 experimental SEs with more than 27,600 conductivity entries. These included 2,900 inorganic SEs, 528 polymer SEs, and 386 gel SEs, corresponding to 824, 99, and 157 unique DOIs, respectively. The dataset also contained 852 computational SEs with 1,223 entries from 201 unique DOIs (**Figure S1**). Representative records with available conductivity–temperature relationships were selected for independent reanalysis.

## Data extraction and reinterpretation

Conductivity values were independently re-extracted from published figures and tables by reconstructing conductivity–temperature relationships using reported axes, scales, and annotations. When necessary, temperature values were converted between °C and K and mapped to Arrhenius coordinates (1000/T) for consistency.

For Arrhenius-type plots, multiple plausible interpretations of the plotted quantity (for example, log(σ), ln(σT), or log(σT)) were considered where axis definitions were incomplete or ambiguous. Extracted values were cross-compared with those reported in the main text, abstracts, or captions. When multiple sources (e.g., literature and curated databases) reported values for the same material, consistency between sources was assessed without assuming any single source to be authoritative.

For cases involving suspected cross-database inconsistencies, values were manually traced from the secondary database back to the original source article. The plotted quantity, axis label, conductivity unit, and required conversions between σT and σ or between S $m^{-1}$, S $cm^{-1}$, and mS $cm^{-1}$ were checked before assigning an issue category.

## Definition and classification of reporting ambiguity

Reporting issues were categorized into four types based on operational criteria:

(i) **Text–figure inconsistency**, defined as a mismatch between reported numerical values and the corresponding graphical representation beyond graphical resolution or expected experimental uncertainty;

(ii) **Annotation ambiguity**, defined as insufficient specification of plotted quantities or axes leading to multiple plausible interpretations;

(iii) **Unit inconsistency**, defined as discrepancies in unit conventions between text, figures, and derived quantities that could alter conductivity values by orders of magnitude;

(iv) **Incomplete measurement context**, defined as missing or unclear information regarding temperature, measurement conditions, or whether values were directly measured, interpolated, or derived from fitting.

Each case was evaluated based on internal consistency within the publication and, where relevant, consistency with downstream database entries. Cases involving more than one issue type were assigned to all applicable categories rather than forced into a single class.

## Case identification and statistical analysis

Identified cases were manually screened and classified according to the criteria above. Representative examples were selected to illustrate each ambiguity type, and the distribution of issue categories was summarized across different electrolyte classes.

For gel polymer electrolytes, the statistics were calculated at the paper level from 83 screened publications. Papers without identifiable reporting issues were counted separately from papers showing text–figure inconsistency or annotation ambiguity.

For inorganic solid electrolytes, the statistics were calculated by issue occurrence rather than by paper, because a single source case could contain multiple issue types. The 16 inorganic source cases listed in **Table S2** correspond to 25 issue occurrences. These issue occurrences were used to construct the inorganic SE bar chart in **Fig. 2D**.

Where ambiguity could not be uniquely resolved, cases were conservatively assigned to the most general applicable category. Cross-comparison between literature-derived values and curated database entries was used to assess the propagation of inconsistencies into downstream datasets.

## Scope and limitations

This analysis does not aim to provide an exhaustive statistical survey of the entire solid electrolyte literature, but rather to identify representative and reproducible types of ambiguity encountered during data curation across multiple data sources. The interpretation of graphical data is inherently dependent on figure resolution and reporting clarity and may involve expert judgment.

Furthermore, curated databases themselves may reflect prior literature inconsistencies, as highlighted by the need for extensive expert validation in recent dataset construction efforts. As such, the identified discrepancies are interpreted as structural characteristics of current reporting practices rather than isolated errors.

**Data availability**

The source papers underlying the representative cases analyzed in this work are listed in Supplementary Tables S1 and S2. The issue classifications used to generate Fig. 2D and the DigBat statistics used to generate Fig. S1 are provided in the Supplementary Information.

**Acknowledgement**

This study was financially supported by the JST Strategic International Collaborative Research Program (SICORP; JPMJSC25E2), the GIMRT Program of the Institute for Materials Research, Tohoku University (Proposal No. 202512-CRKKE-0212), the AY2024/2025 TUMUG Support Program, and the Basic Research Grant from the TEPCO Memorial Foundation. The authors acknowledge the use of MASAMUNE-IMR at the Center for Computational Materials Science, Institute for Materials Research, Tohoku University (No. 202512-SCKXX-0205).

## Supplementary Information

# When Literature Data Mislead Artificial Intelligence in Materials Discovery

Qian Wang[1], Ying Li[1], Ryuhei Sato[2], Hidemi Kato[3], Shin-ichi Orimo[1,3], Hao Li[1], Eric Jianfeng Cheng[1,*]

[1] Advanced Institute for Materials Research (WPI-AIMR), Tohoku University, Sendai 980–8577, Japan

[2] Department of Materials Engineering, The University of Tokyo, Tokyo 113-8656, Japan

[3] Institute for Materials Research, Tohoku University, Sendai 980–8577, Japan

* Email: ericonium@tohoku.ac.jp

**Table S1**. Gel SE papers with issues

| Case ID | DOI | Issue category |
|---|---|---|
| **G1** | 10.1002/smll.202504201 | Text–figure inconsistency |
| **G2** | 10.1016/j.ensm.2025.104469 | Text–figure inconsistency |
| **G3** | 10.1016/j.cej.2024.149757 | Text–figure inconsistency |
| **G4** | 10.1016/j.cej.2024.151161 | Text–figure inconsistency |
| **G5** | 10.1002/aenm.202402362 | Text–figure inconsistency |
| **G6** | 10.1016/j.jpowsour.2024.234262 | Text–figure inconsistency |
| **G7** | 10.1063/5.0134474 | Text–figure inconsistency |
| **G8** | 10.1039/d2cc04128f | Text–figure inconsistency |
| **G9** | 10.1002/aenm.202500887 | Text–figure inconsistency |
| **G10** | 10.1021/acsaem.4c00375 | Annotation ambiguity |
| **G11** | 10.1002/adfm.202514477 | Annotation ambiguity |
| **G12** | 10.1002/batt.202400463 | Annotation ambiguity |

**Table S2**. Inorganic SE cases with issues

| **Case ID** | **DOI** | **Issue category** |
|---|---|---|
| **I1** | 10.1021/acs.nanolett.4c05460 | Annotation ambiguity + Unit inconsistency |
| **I2** | 10.1021/acs.nanolett.4c03750 | Unit inconsistency |
| **I3** | 10.1016/j.nanoen.2021.106674 | Annotation ambiguity + Unit inconsistency |
| **I4** | 10.1039/d5ta01706h | Unit inconsistency |
| **I5** | 10.1002/adma.202512961 | Annotation ambiguity + Unit inconsistency |
| **I6** | 10.1016/j.cej.2025.164300 | Unit inconsistency |
| **I7** | 10.1016/j.cej.2025.162128 | Annotation ambiguity + Unit inconsistency |
| **I8** | 10.1016/j.ssi.2023.116244 | Unit inconsistency + Text–figure inconsistency |
| **I9** | 10.1021/acsaem.5c00020 | Unit inconsistency |
| **I10** | 10.1016/j.mtener.2025.102077 | Unit inconsistency + Text–figure inconsistency |
| **I11** | 10.1039/C9CP05329H | Annotation ambiguity + Text–figure inconsistency |
| **I12** | 10.1039/C9SE01162E | Annotation ambiguity |
| **I13** | 10.1039/d0cp03442h | Text–figure inconsistency / Incomplete measurement context |
| **I14** | 10.1016/j.jallcom.2025.180413 | Unit inconsistency |
| **I15** | 10.1002/ejic.202300382 | Unit inconsistency |
| **I16** | 10.1016/S0167-2738(98)00414-7 | Annotation ambiguity + Unit inconsistency |

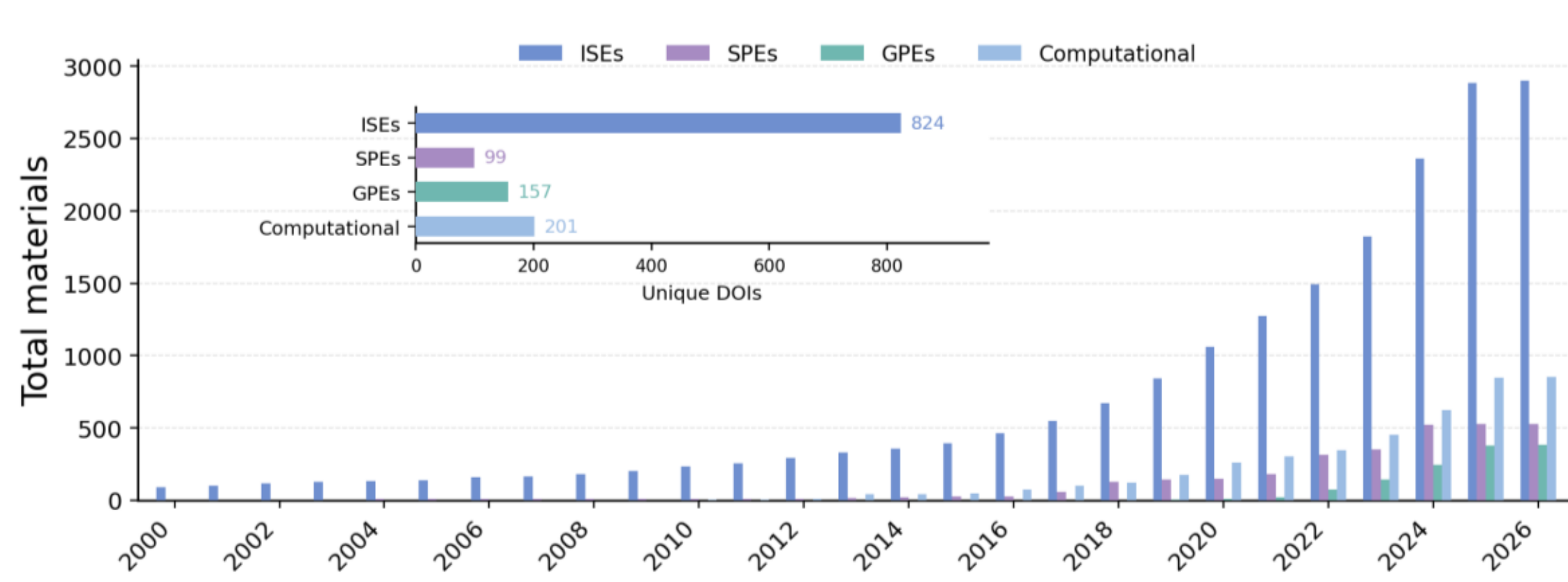

**Figure S1**. Statistical overview of solid electrolyte data in the DigBat database. Temporal evolution of the total number of materials recorded in DigBat, categorized into inorganic solid electrolytes, solid polymer electrolytes, gel polymer electrolytes, and computational data. The inset summarizes the number of unique DOIs associated with each category. All statistics are derived from the DigBat dataset.